\documentclass[%
 reprint,
showpacs,preprintnumbers,
 amsmath,amssymb,
 prl,
]{revtex4-1}

\usepackage{graphicx}
\usepackage{dcolumn}
\usepackage{bm}
\usepackage{textgreek}

\begin{document}


\title{Controlling Coherent Quantum Dot Interactions}

\author{Eric W. Martin}
\author{Steven T. Cundiff}%
 \email{cundiff@umich.edu}
\affiliation{%
 Applied Physics Program and Department of Physics\, University of Michigan, Ann Arbor, MI 48109-1040, USA\\
}%

\date{\today}

\begin{abstract}
We measure many-body interactions in isolated quantum dot states using double-quantum multidimensional coherent spectroscopy. Few states are probed in a diffraction limited spot, which is enabled by a novel collinear scheme in which radiated four-wave-mixing signals are measured with phase resolution. Many-body interactions are enhanced by an additional prepulse tuned to the delocalized quasi-continuum states. We propose this effect as a method for controlling coupling between quantum states.
\end{abstract}

\maketitle



Quantum dots (QDs) are often described as being non-interacting artificial atoms. Some optical spectroscopic experiments have been used to conclude that there are no measurable many-body interactions present for resonant excitation of interfacial ODs, which would support treating these QDs as non-interacting \cite{Bonadeo1998}. However, other optical techniques have yielded signatures of interactions between these QDs \cite{Langbein2011,Fan1998,Moody2011}. 
Outside of the spectroscopic differences, discrepancies exist regarding the presence of many-body effects in QD lasers \cite{Chow2013}. The benefits of QD lasers arise from the discrete and narrow energy levels of QDs, but they are usually pumped by the injection of delocalized carriers \cite{Bimberg1997}. Since many-body effects play a tremendous role in the theoretical treatment of semiconductors \cite{Chow1999}, it is important to understand the relevant interactions for calculating QD laser properties.

Excitons and trions confined to QDs are potential candidates for qubits in quantum information \cite{Press2008,Berezovsky2008,Wu2006}. 
The electronic states of a QD are accessible both optically and electronically. Also, the high oscillator strengths of electronic transitions in solid state systems facilitate their measurement and manipulation. Coherent control with ultrafast Rabi rotations has been demonstrated on both single and ensemble QD systems \cite{Li2003,TakeshiPRL}. However, controlled qubit interaction remains one of the most challenging requirements for a functional quantum computer with few implementations for spin states in QDs \cite{Kim2011,Spatzek2011} and none for the electronic states. The localization of excitons in QDs that gives them the benefit of being difficult to decohere also makes them difficult to entangle, or couple \cite{Serbyn2013}.

Here we observe that the excitation of delocalized states not only enhances many-body effects, in agreement with theory \cite{Schneider2004}, but can also turn them on. The physical mechanism responsible for enhancing many-body interactions in QDs may explain discrepancies in the literature. The mechanism may also be applied for turning on electronic coupling between isolated QD states.

We use ultrafast coherent spectroscopy techniques to directly probe coupling and many-body interactions in a sub-micron-sized region containing a small number of distinct epitaxially-grown GaAs interfacial QDs at a temperature of 6 K. These interfacial QDs are exciton states bound by monolayer fluctuations in a narrow 4.2 nm GaAs quantum well with Al$_{0.3}$Ga$_{0.7}$As barriers \cite{Gammon1996}. The decreased transverse confinement binds the localized excitons by 10 meV, which energetically separates them from the delocalized quantum well resonances. Because of the large spatial separation (averaging 300-400 nm) between QDs, the natural coupling between them is minimal. By resonantly exciting higher energy delocalized exciton states in the quantum well we open coherent coupling channels between localized excitons. After pulsed excitation of the delocalized states, we use double-quantum spectroscopy to directly measure exciton-exciton interactions between isolated single quantum systems.

To probe the localized QD response to resonant excitation of delocalized states, we use multidimensional coherent spectroscopy (MDCS). MDCS is a transient four-wave mixing spectroscopy that has evolved from four-wave mixing techniques responsible for realizing the importance of considering Coulomb interaction effects in semiconductor quantum wells \cite{Kim1992,Weiss1992}. In MDCS, the phase-resolved evolution of the nonlinear response is measured as a function of the evolution of a phase-resolved linear response. These measurements result in spectra with two or more dimensions that correlate absorption, emission, and evolution energies of sample coherences \cite{Cundiff2013}. There are various pulse sequences we can use to measure coherent processes. Single-quantum pulse sequences developed for MDCS are used to directly measure coupling between QD states and the intrinsic linewidth of the QD resonances. A double-quantum pulse sequence directly measures signals resulting from many-body interactions.

Most MDCS techniques rely on k-vector selection, which requires a finite spot. With few exceptions \cite{Tekavec2007,Nardin2013,Langbein2011}, these techniques are thus limited to the study of spatially extended states or dense ensembles. We have developed a variant of collinear techniques \cite{Nardin2013,Tekavec2007} that instead uses heterodyne detection to measure radiated MDCS signals. To distinguish the optical signal from the co-propagating excitation beams, each beam is tagged with a different radio frequency shift using acousto-optic modulators. The radiated third-order nonlinear signals are shifted by radio frequencies that depend on the excitation beams used to generate them. The interference between the radiated nonlinear signal and a separately tagged local oscillator (LO) has a beat note at the difference between their frequency tags. A lock-in amplifier measures the signal at the phase-matched modulation frequency. We accurately measure the signal phase by co-propagating all beams with a continuous-wave (CW) laser that samples all of the mechanical fluctuations that contribute to phase noise. We interfere these CW beams with each other on two detectors and use those measurements to calculate a phase-corrected reference at the signal frequency. Since the reference is affected by the same path-length fluctuations as the signal, the measured signal has a meaningful phase with respect to the excitation pulses. See the supplemental material for details about the experimental configuration \cite{Supplement}.

\begin{figure}[htp!]
  \includegraphics[width=0.34\textwidth]{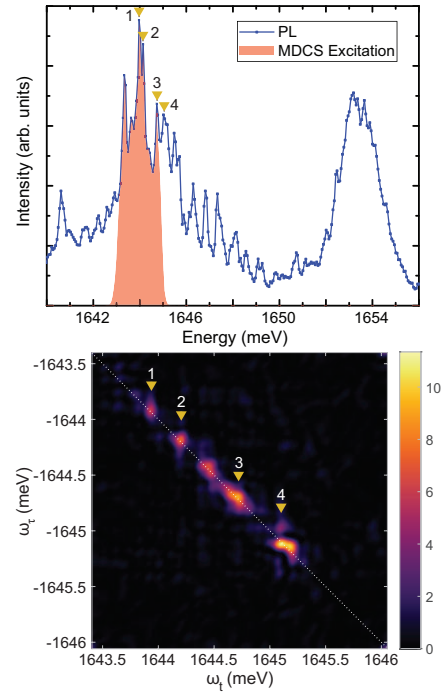}
  \caption{Top: Photoluminescence (PL) excited by a 633 nm laser is measured on a spectrometer with 100 \textmu eV resolution. Features below 1650 meV are attributed to localized quantum dot states that we spatially isolated with a diffraction limited 700 nm spot. The wide feature above 1650 meV are the residual two-dimensional (quantum well) states. The spectral region measured by multidimensional coherent spectroscopy (MDCS) in this paper, shaded in red, is determined by the shaped laser spectrum we use. Bottom: Single-quantum MDCS spectrum of the same region allows for comparison of the oscillator strengths of resonances and reveals that some of the weakly excited higher energy states have very high oscillator strengths.}
  \label{fig:PLSI}
\end{figure}

The MDCS pulses are spectrally tuned and filtered to resonantly excite only the localized exciton states. In Figure~\ref{fig:PLSI} we compare a single-quantum nonlinear MDCS measurement to the photoluminescence spectrum from the same sample region. These spectra do not match because nonlinear emission and radiative recombination for a resonance have different dependences on each QD's dipole moment. In this case, however, the strong localized state resonances, labeled 1-4, emit at the same center frequencies with mostly comparable strengths. Exceptions exist at higher energies, and we show that resonance 4 has an exceptionally high nonlinear response relative to the lower energy resonances. Since these higher energy states are generally more delocalized, we attribute this enhanced nonlinearity to many-body effects, which are known to be the dominant source of nonlinear optical responses in semiconductor quantum wells \cite{Kim1992,Weiss1992,Chemla2001}.

Using a rephasing pulse sequence, which is typically used in ensemble MDCS measurements to separate inhomogeneous and homogeneous broadening such that homogeneous linewidths can be measured \cite{Siemens2010}, we measure an average low temperature QD linewidth between 27 and 28 \textmu eV. This measurement is in agreement with previous low excitation density experiments \cite{Hess1994,Gammon1996,Fan1998}. At high excitation densities there has been some disagreement in linewidth measurements found in the literature. Four-wave mixing measurements of interfacial QD ensembles have observed large dephasing rates at high densities, a feature that would make these QDs resemble higher dimensional systems \cite{Fan1998,Moody2011}. However, linewidth measurements of interfacial QDs with high enough spatial resolution to distinguish the QDs do not depend on excitation density \cite{Bonadeo1998}. Taking aspects from all these experiments to understand the source of the disagreement, we use spectrally narrowed pulsed light that only excites localized states and a small excitation spot. We measure that the low temperature linewidth is independent of excitation density and conclude that a likely source of dephasing in experiments with large spot sizes is sample heating.

\begin{figure*}[htp!]
  \includegraphics[width=\textwidth]{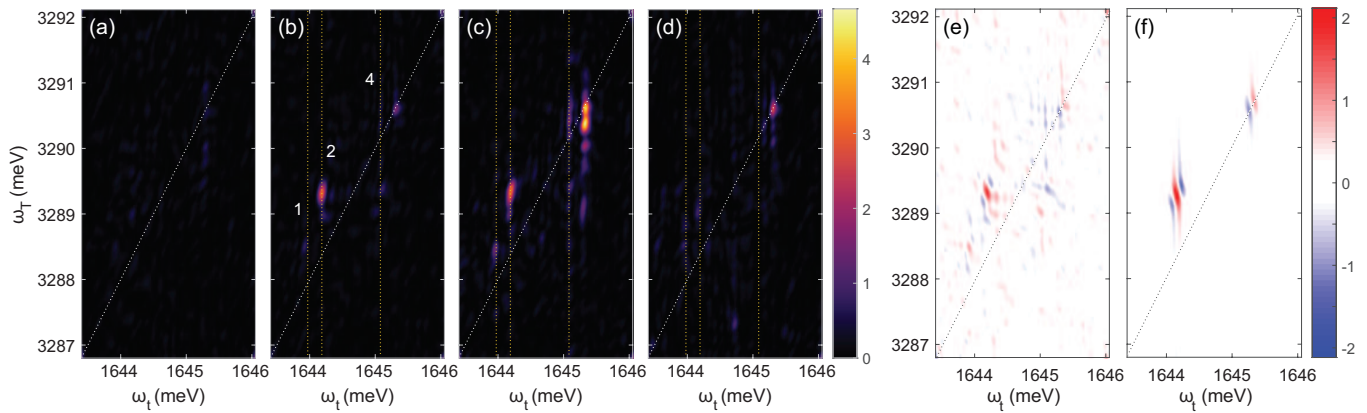}
  \caption{Double-quantum MDCS spectra as a function of prepulse power for (a) 0, (b) 500, (c) 1500, and (d) 4000 photons per pulse. A coupling feature between QDs 2 and 4 appears in (b), which corresponds to a many-body interaction that has been turned on between those resonances. A new higher energy feature grows to a maximum on the diagonal at 1645.3 meV in (c). In (d) the prepulse has saturated the system so the coherent signal is significantly degraded. The real part of (b) is plotted in (e), and a simulation of these features is plotted in (f). The simulation is used to determine the dominant many-body terms that give rise to each of the double-quantum signals.}
  \label{fig:SIII}
\end{figure*}

In order to observe a double-quantum MDCS signal, it is necessary that two interacting excited states coherently evolve simultaneously. With resonant excitation of only the QD states, these signals can result from three interactions. A double-quantum signal resulting from 1) biexcitons in non-interacting self-assembled QDs has been measured \cite{Kasprzak2017}, but excitation of these signals requires enough bandwidth to excite both the exciton and biexciton. Our sample has been well characterized using MDCS, and it is known that the biexciton binding energy of an ensemble of these dots increases with emission energy from 3.3 to 3.8 meV, and it has a distribution about that center binding energy of less than 270 \textmu eV \cite{Moody2013}. The distribution of biexciton binding energies for a set of individually measured quantum dots, which has the advantage over ensemble measurements of being able to exactly correlate biexciton and exciton emissions, is just 200 \textmu eV \cite{Kasprzak2012}. The MDCS beams have a narrow bandwidth of 2 meV with sharp spectral edges (0.2 meV) such that we cannot doubly excite a single QD (more details in the Supplemental Material \cite{Supplement}). Thus, the only source of a signal from a QD resonance can be 2) interaction between two different QDs.  The interactions in both measurements, however, are very weak and require that the QDs have a very close proximity. 3) If a weakly localized state is large enough for it to be doubly excited without forming a bound state, the resulting double-quantum signal would more closely resemble those measured in quantum wells \cite{Karaiskaj2010}. Using the phase of the double-quantum signal, we can distinguish binding and scattering many-body interactions \cite{Mukamel2008}, so we can identify the above sources of double-quantum signals. By spatially isolating just a few quantum dots within a 700 nm focus, we can thus directly measure interactions at the single excitation level.

We use double-quantum MDCS to determine if QD states produce interaction induced signals. 
On-diagonal signals in a double-quantum MDCS spectrum correspond to self interaction, which we attribute to either spatially large localization sites that confine multiple degenerate excitations or spatially adjacent nearly degenerate quantum dot sites. Off-diagonal signals are due to many-body interaction between two excitations at different energies. For example interactions between two frequencies $\omega_1$ and $\omega_2$, these signals can emit at either of those frequencies and will evolve at their sum: $\omega_T = \omega_1 + \omega_2$. These weak off-diagonal signals most likely result from radiative interaction between adjacent quantum dot states, which has been shown to have a long range exceeding 400 nm \cite{Langbein2011}. Though weak, we measure interactions between few resonantly excited QDs over the sample, and weak interactions between resolved QD states have very recently been measured in self-assembled QDs \cite{Kasprzak2017}.

In order to measure the effect of delocalized quantum well excitations on QD interactions we excite the delocalized quantum well states with a pre-pulse that is spectrally filtered to excite only the quantum well states. The pre-pulse has a power between 10 and 80 nW (500-4000 photons per pulse), and it arrives 20 ps before the first MDCS pulse so that only the incoherent population it creates is present when the MDCS spectrum is measured. As shown by comparison of Figs.~\ref{fig:SIII}(a) and (b) a small excitation of the delocalized states greatly enhances interaction among localized QD excitons, which results in a strongly enhanced off-diagonal peak in the double-quantum MDCS spectrum. From the evolution and emission energies it is evident that this feature corresponds to coupling between resonances 2 and 4 labeled in Fig.~\ref{fig:PLSI}. As the prepulse power is increased in Fig.~\ref{fig:SIII}(c), lower energy QD states are filled due to dynamic localization of the extended states created by the prepulse, and higher energy double-quantum features are enhanced. The strong on-diagonal feature at $\omega_{t} = 1645.3$ meV does not strongly show up in single-quantum MDCS without some prepulse excitation either, shown in the supplementary material \cite{Supplement}. Along with the enhanced oscillator strength of the high energy features measured with single-quantum MDCS, the presence of this state on the diagonal is evidence that it is a higher dimensional state than a QD since it can be doubly excited. With a much higher prepulse excitation in Fig.~\ref{fig:SIII}(d), all double-quantum coherences are blocked by filling of the QD states.

In Fig.~\ref{fig:SIII}(e) we plot the real part of (b). To interpret the phase of the double-quantum MDCS signal requires a simple simulation in which we consider the phase of the linear responses to each pulse. We simulate the nonlinear response by analytically solving a perturbative expansion of the density matrix for two coupled two-level systems \cite{Supplement,hamm2011concepts}. The energy level scheme consists of a ground state, a single-excited state for each QD, and double-excited state representing simultaneous excitation of both QDs. Many-body interactions break the symmetry between the transition into the singly excited and doubly excited states, which is represented by a shift or broadening of the doubly excited level. By accurately measuring the phase of the double-quantum signal, we can identify the many-body terms that give rise to those signals. In order to produce an accurate simulation of the data in Fig.~\ref{fig:SIII}(f), we find that the coupling feature corresponding to the interaction of QDs 2 and 4 is an excitation induced red shift of the doubly excited state. A red shift of the doubly excited state is indicative of a weak binding between the two states \cite{Mukamel2008}. The on-diagonal feature, on the other hand, results from a combination of excitation induced dephasing and blue shift. These exciton scattering effects are typically measured in quantum wells, further supporting that this higher energy state is higher dimensional than a QD. We find similar results for QD states in other regions of the sample. 

\begin{figure}[htp!]
  \includegraphics[width=0.4\textwidth]{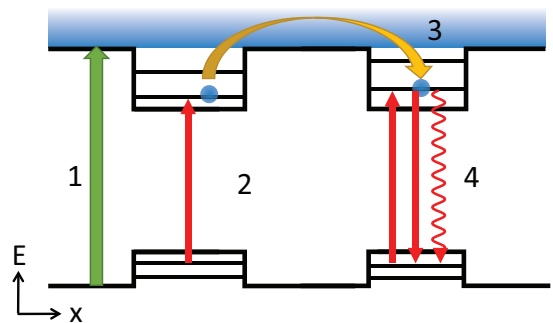}
  \caption{Schematic of pulse sequence applied to spatially isolated interfacial quantum dots. Interactions between QDs are very weak, but we can turn on coupling by creating delocalized quantum well carriers with (1) a higher energy prepulse. We probe the induced interactions with (2) two coherent pulses that create a double coherence of different excitonic transitions. (3) The interaction between the coherences is mediated by the quantum well carriers, and (4) we read out the interaction with a coherent third pulse that begins emission of a coherent four-wave-mixing signal. Though the prepulse also creates incoherent excitations of the QD states, this is negligible for low prepulse powers and only serves to degrade the overall signal.}
  \label{fig:schem}
\end{figure}

The prepulse enhancement of many-body interactions between QD states is illustrated in Figure~\ref{fig:schem}. The delocalized carriers in the quasi-continuum states serve to mediate interactions between spatially separated QD states. The QD excitations are localized to roughly 50 nm islands, but delocalized excitons are much more extended. So while there is no wave function overlap of individual QDs, the wave function of the quantum well excitation introduces coupling of localized states. The enhanced range of interaction between QDs is still limited by the finite mobility of the delocalized excitons, roughly 15 cm$^2$/s in a thin quantum well \cite{Hegarty1985}. Therefore only a few of the localized excitations within a given spot will be within the range of each other to interact via the delocalized excitons.

Existing microscopic theory supports the concept that excitation induced dephasing and shift in interfacial QDs arises from interactions with quasi-continuum quantum-well states \cite{Schneider2004}. Schneider \emph{et al.} discuss broadening and redshift that is dependent on density, effects they determine by calculating the renormalized electronic states. They also discuss that their calculation of density dependent dephasing in interfacial QDs is equivalently relevant to self-assembled quantum dots electronically coupled to a wetting layer.


Though we have presented a method for turning on coupling between isolated interfacial QDs, this method may be generalized to coupling any localized quantum states in physical contact with higher energy delocalized states; at least states that may be excited in a controlled way. We see immediate benefit in the ability to control coupling self-assembled QDs in contact with the higher energy wetting layer. Also, in light of recent findings of long-lived localized states in transition metal dichalcogenides (TMDCs) \cite{Srivastava2015,Koperski2015,He2015,Chakraborty2015,Tonndorf2015}, this work could be applied to these states which could be coupled through the highly delocalized TMDC exciton states.

Another major outlook for measuring physical systems at the nanoscale is that the coupling of individual QDs to delocalized excitons introduces a new method for studying the delocalized states. The locations of QDs can be determined with much higher accuracy than the optical resolution. Since QDs are spectrally distinct, one could thus consider using measurements of QDs separated by known distances to probe length scales and transport in the continuum states with the resolution of a QD.

In summary, we have developed a collinear MDCS technique that utilizes dynamic phase cycling to probe nonlinear responses with high sensitivity and phase resolution. We have used this technique at the diffraction limit to resolve individual QD oscillators. We demonstrated both double-quantum and single-quantum measurements, and with this technique it is actually simple to selectively measure even higher order nonlinear expansion terms. Using double-quantum MDCS, which is sensitive only to many-body effects, we measure an absence of many-body effects in interfacial quantum dots with resonant excitation. However, we find that these effects can be enhanced by excitation of the delocalized quantum-well states using a prepulse.
This work helps to explain some discrepancies in the literature in which weak excitation of continuum states with broadband pulses has not been explicitly considered. From an applications standpoint, we present this prepulse technique as a way of turning on coupling between quantum states.

We thank D. Gammon and A. Bracker at the Naval Research Laboratory for the interfacial quantum dot sample, and we thank C. Smallwood for useful feedback and discussions. This work is supported by the National Science Foundation (NSF) Division of Materials Research (DMR).



\bibliography{naturalQD}
\end{document}